\documentclass[smus]{snow2e}
\usepackage{graphics,epsfig,rotating}

\begin{document}

\title{Search Limits for Extra Neutral Gauge Bosons at High Energy 
Lepton Colliders\thanks{This research 
was supported in part by the Natural Sciences and Engineering 
Research Council of Canada.}}

\author{Stephen Godfrey \\ 
{\it Ottawa-Carleton Institute for Physics} \\
{\it Department of Physics, Carleton University, Ottawa CANADA, K1S 5B6} }

\maketitle

%% Get rid of page numbering
\thispagestyle{empty}\pagestyle{empty}

\begin{abstract} 
We study and compare the discovery potential 
for heavy neutral gauge bosons ($Z'$) at the various $e^+e^-$ and 
$\mu^+\mu^-$ colliders that have been 
proposed.  Typical search limits for the $e^+e^-$ colliders 
are $2-10\times \sqrt{s}$ 
with the large variation reflecting the model dependence of the limits.
The search limits for the $\mu^+\mu^-$ colliders are slightly lower.
Polarization and flavour tagging are important in realizing 
the highest discovery limits possible. Because the 
search limits are
based on indirect inferences of deviations from standard 
model predictions, they are sensitive to systematic errors.

\end{abstract}

\section{Introduction}
Extended gauge symmetries and the associated heavy neutral gauge 
bosons, $Z'$, are a feature of many extensions of the standard model 
such as grand unified theories, Left-Right symmetric models, and 
superstring theories.  If a $Z'$ were discovered it would have 
important implications for what lies beyond the standard model.  
It is therefore important to study and compare the discovery 
reach for extra gauge bosons at the various facilities 
that are under consideration for the future. Included in the list of 
proposed facilities considered at the Snowmass'96
workshop are high energy $e^+e^-$ and $\mu^+ \mu^-$ colliders. 
In this report we update previous studies 
\cite{cvetic,godfrey,capstick,hewett,hewett2}
to include these high energy lepton colliders.

\section{Models}

Quite a few models predicting extra gauge bosons exist in the 
literature.  We will present 
search limits for several of these models which, although far 
from exhaustive, form a  representative set for the purposes 
of comparison.  The models chosen for study are listed below but for 
details we direct the interested reader to ref. \cite{godfrey} and 
references therein.

\begin{description}
\item[(i)] Effective rank-5 models originating from $E_6$ grand unified 
theories are conveniently labelled in terms 
of the decay chain $E_6 \to SO(10) \times U(1)_\psi \to SU(5)\times 
U(1)_\chi \times U(1)_\psi \to SM \times U(1)_{\theta_{E_6}}$.  
Thus, the $Z'$ charges are given by linear combinations of the 
$U(1)_\chi$ and $U(1)_\psi$ charges resulting in the $Z'$-fermion 
couplings:
\begin{equation}
g_{Z'} (Q_\chi \cos \theta_{E_6} + Q_\psi \sin\theta_{E_6})
\end{equation}
where $\theta_{E_6}$ is a free parameter which lies in the range 
$-90^\circ \leq \theta_{E_6} \leq 90^\circ$.
Specific models of interest are model $\chi$ ($\theta_{E_6}=0^\circ$)
corresponding to the extra $Z'$ of $SO(10)$, model $\psi$ 
($\theta_{E_6}=90^\circ$) corresponding to the extra $Z'$ of $E_6$, 
and model $\eta$ ($\theta_{E_6}=\arctan -\sqrt{5/3}$) 
corresponding to the extra $Z'$ arising in some superstring theories.

\item[(ii)] The Left-Right symmetric model (LRM) extends the standard model 
gauge group to $SU(2)_L \times SU(2)_R \times U(1)$.
It has a parameter $\kappa$ 
with $0.55 \leq \kappa^2 \equiv (g_R/g_L)^2 \leq 1-2$.
We assume $\kappa=1$ in our analysis which 
corresponds to strict left-right symmetry. 

\item[(iii)] The Alternative Left-Right Symmetric model (ALRM)
originates from $E_6$ GUT's and is also based on the electroweak 
gauge group $SU(2)_L \times SU(2)_R \times U(1)$.  Here the 
assignments for $T_{3L(R)}$ differ from that of the usual LRM 
due to an ambiguity in how the fermions are embedded in the 27 
representation of $E_6$.

\item[(iv)] The ``sequential'' Standard Model (SSM) consists of a $Z'$ 
which is just a heavy version of the SM $Z^0$ boson with identical 
couplings. Although it is not a realistic model 
it is often used as a benchmark and for purposes of comparison.

\item[(v)] The Harvard Model (HARV) is based on the gauge 
group $SU(2)_l \times SU(2)_q \times U(1)_Y$, i.e., left-handed 
leptons (quarks) transform as doublets under $SU(2)_l$ ($SU(2)_q$)
and singlets under  $SU(2)_q$ ($SU(2)_l$), and right-handed fields 
are singlets under both groups.  This model has a parameter
$\phi$ which lies in 
the range $0.22 \leq \sin\phi \leq 0.99$.  We take $\sin\phi=0.5$ in 
our calculations.  The $Z'$ is purely left 
handed in this model.

\end{description}

There are numerous other models in the literature predicting $Z'$'s 
but the subset described 
above have properties reasonably representative of broad classes of 
models,  at least for the purposes of comparing search
limits of high energy colliders.

\section{Calculations and Results}

At $e^+e^-$ colliders searches for $Z'$'s are indirect, being inferred 
from deviations from the standard model predictions due to 
interference between the $Z'$ propagator and the $\gamma$ and $Z^0$ 
propagators \cite{e+e-}.  
This is similar to PEP/PETRA seeing the standard model 
$Z^0$ as deviations from the predictions of QED.
The basic process is $e^+e^-_\lambda \to f\bar{f}$ where $f$ 
could be leptons $(e,\; \mu ,\; \tau)$ or quarks $(u, \; d, \; c,\; 
s,\; b, \; t)$ and $\lambda$ denotes the $e^-$ polarization.  
The cross section for the basic process is given by:
\begin{equation}
{{d\sigma_L} \over {d\cos\theta}} = {{\pi \alpha^2}\over{4s}}
\left\{ |C_{LL}|^2(1+\cos\theta)^2 +|C_{LR}|^2 (1-\cos\theta)^2 \right\}
\end{equation}
where
\begin{eqnarray}
C_{ij} & =&  -Q_f 
+{{C_i^e C_j^f}\over {c_w^2 s_w^2}} 
{s\over{(s-M_Z)^2+i\Gamma_Z M_Z}} \nonumber \\
& &  +{{(g_{Z'}/g_{Z^0})^2 {C_i^e}' {C_j^f}'}\over {c_w^2 s_w^2}} 
{s\over{(s-M_{Z'})^2+i\Gamma_{Z'} M_{Z'}}} 
\end{eqnarray}
where the $C_i^f$ are the SM $Z^0$ couplings and the ${C_i^f}'$ are the $Z'$ 
couplings.
For right-handed electrons make the substitutions $C_{LL} \to C_{RR}$ 
and $C_{LR} \to C_{RL}$.
From these expressions we can obtain everything else we need.  
For example, the total differential cross section is just
\begin{equation}
{{d\sigma} \over {d\cos\theta}} ={1\over 2} 
\left[{{d\sigma}_L \over {d\cos\theta}}+{{d\sigma}_R \over {d\cos\theta}}
\right]
\end{equation}
and the various polarized or unpolarized total cross sections can be 
obtained by integrating these expressions.  The spin averaged 
(unpolarized) cross section is given by:
\begin{equation}
\sigma = {{\pi \alpha^2}\over {3s}} [ |C_{LL}|^2 + |C_{RL}|^2
+ |C_{RL}|^2 + |C_{RR}|^2 ]
\end{equation}

From the basic reactions a number of observables can be 
used to search for the effects of $Z'$'s: 
\begin{itemize}
\item $\sigma^f$ --- the cross section to specific final state fermions
\item $R^{had}= \sigma^{had}/\sigma_0$ ---
the ratio of the hadronic to the QED point cross section
\item $A^f_{FB}$ --- the forward-backward asymmetry to fermion $f$ 
in the final state;
\begin{equation}
A_{FB}=  
{ { \left[{ \int^1_0 -\int^0_{-1} }\right] d\cos\theta 
{{d\sigma}\over{d\cos\theta}} } \over
{ \left[{ \int^1_0 -\int^0_{-1} }\right] d\cos\theta 
{{d\sigma}\over{d\cos\theta}} } }
\end{equation}
\item $A^f_{LR}$ --- the left-right asymmetry with fermion $f$ in the final 
state;
\begin{equation}
A_{LR}^f = {{\sigma(e^-_L)-\sigma(e^-_R)}\over
{\sigma(e^-_L)-\sigma(e^-_R)}}
\end{equation}
\item $A^{had}_{LR}$ --- the left-right asymmetry with 
hadrons in the final state
\item $A^f_{FB}(pol)$ --- the polarized forward-backward double 
asymmetry with fermion $f$ in the final state;
\begin{equation}
A_{FB}^f(pol) = {{(\sigma_L^F-\sigma_R^F)-(\sigma_L^B-\sigma_R^B)}
\over {(\sigma_L^F-\sigma_R^F)+(\sigma_L^B-\sigma_R^B)}}
\end{equation}
\item $P_\tau$ --- the tau polarization
\end{itemize}
In these expressions we generally  take the index 
$f = \; \mu, \; \tau, \; c, \;b$ and $had=$`sum over all hadrons' to 
indicate the final state fermions.

To obtain discovery limits for new physics, which in this case means evidence 
for extra neutral gauge bosons, we look for statistically significant 
deviations from standard model expectations.  In Fig. 1 a number of 
observables are shown with their standard model values and for 
various $Z'$'s as a function of the $Z'$ mass.  The error bars are 
based on the statistics expected in the standard model. 
(Note that there is a cheat here in that I did not include $c$ and 
$b$-quark tagging efficiences in the statistical errors so that in 
reality some of the error bars should really be bigger than what is shown.)
What is important to note here is that the different observables have 
different sensitivities to the different models.  For example, of the 
models shown, 
$\sigma(e^+e^-\to \mu^+ \mu^-)$ is most sensitive to $Z_{ALR}$ 
while $R^{had}$ is most sensitive to $Z_\chi$.  Similarly, 
$A_{FB}^c(pol)$ is most sensitive to $Z_{ALR}$ while $A_{FB}^b(pol)$
is most sensitive to $Z_\chi$.  Therefore to have the highest possible 
reach for the largest number of possible models
it is important to include all possible observables.

\begin{figure}[t]
\leavevmode
\centerline{\epsfig{file=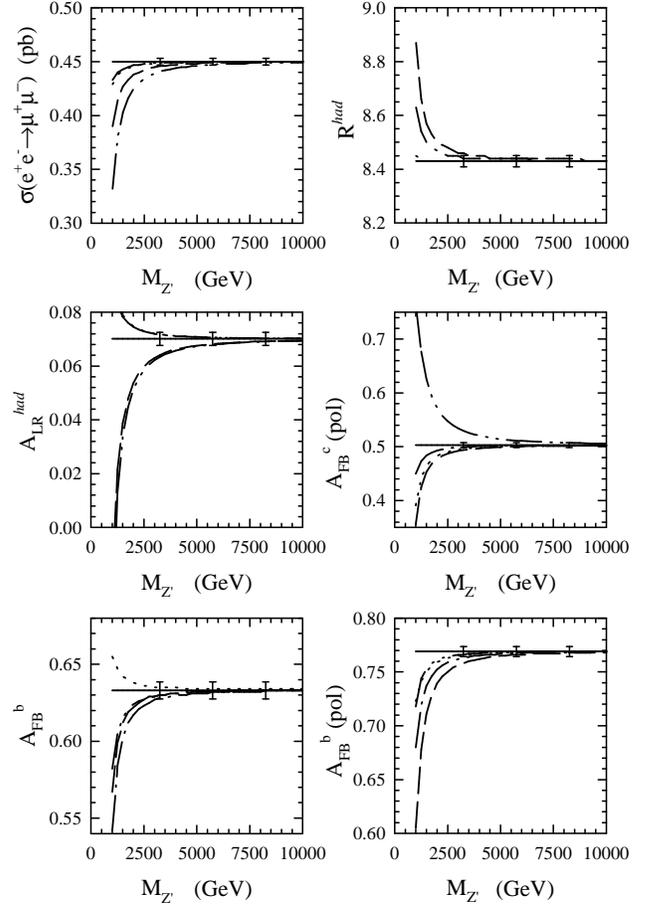,width=9.7cm,clip=}}
\caption{Some $e^+e^-$ observables showing their standard model values 
and their values for a $Z'$ as a function of $M_{Z'}$ for $\sqrt{s}=500$~GeV.  
The solid line is the 
standard model value, the dashed line is for $Z_\chi$, the dotted line 
for $Z_\eta$, the dot-dashed line for $Z_{LR}$ and the dot-dot-dash 
line for $Z_{ALR}$.
The error bars are based on the statistical error assuming an integrated
luminosity of 50~fb$^{-1}$.}
\end{figure}

We quantify the sensitivity to an extra gauge boson by comparing the 
predictions for various observables assuming the presence of a $Z'$ to 
the predictions of the standard model 
and constructing the $\chi^2$ figure of merit;
\begin{equation}
\chi^2 = \sum_i \left( { {  {\cal O}^{Z'}_i - {\cal O}^{SM}_i } \over 
{\delta {\cal O}^{SM}_i  } } \right)^2
\end{equation}
where the sum is over observables included in the $\chi^2$ and 
$\delta{\cal O}$ is the experimental error of the observable.
The contributions of some observables to the $\chi^2$ for $Z_\chi$, 
$Z_\eta$, $Z_{LR}$, and $Z_{ALR}$ are shown in figure 2.  
There are several points to note from this figure.  First, we again 
see that the observables have different sensitivities to the various 
models so that it is important to consider as many observables as 
possible.  Second,  the polarization asymmetries are in many cases the 
most sensitive observables so that polarization is potentially very 
important for searches for $Z'$'s.  And finally, flavour tagging can 
contribute a considerable amount of information.

\begin{figure}[t]
\leavevmode
\centerline{\epsfig{file=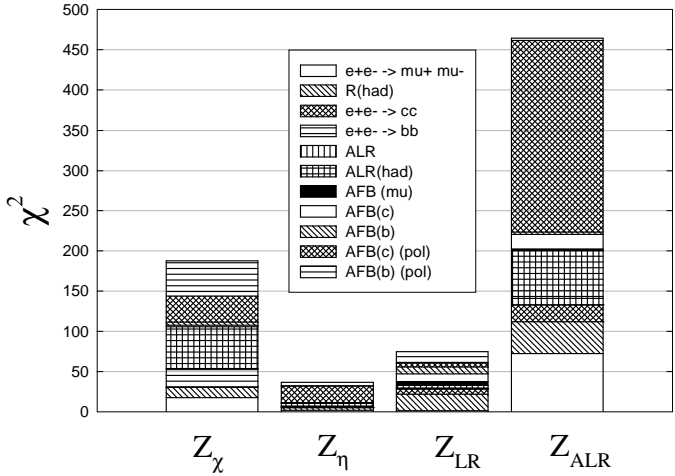,width=9.5cm,clip=}}
\caption{The contributions to $\chi^2$ for the observables
$\sigma(e^+e^- \to \mu^+\mu^-)$, $R^{had}$, $\sigma(e^+e^- \to 
c\bar{c})$,
$\sigma(e^+e^- \to b\bar{b})$, $A_{LR}^\mu$, $A_{LR}^{had}$, $A_{FB}^c$,
$A_{FB}^b$, $A_{FB}^c(pol)$, and $A_{FB}^b(pol)$ for models
$Z_\chi$, $Z_\eta$, $Z_{LR}$, and $Z_{ALR}$.  These are based on 
$\sqrt{s}=500$~GeV, L=50~fb$^{-1}$, and $M_{Z'}=2$~TeV.  The $\chi^2$ 
is based solely on the statistical error.  We assume 100\% polarization 
and do not include finite $c$ and $b$-quark detection efficiencies.}
\end{figure}

To obtain ``discovery'' limits for the $e^+e^-$ case we 
include the eighteen observables: $\sigma^\mu$, $\sigma^\tau$, 
$\sigma^c$, $\sigma^b$, $R^{had}$, $A_{FB}^\mu$, $A_{FB}^\tau$, 
$A_{FB}^c$, $A_{FB}^b$, $A_{LR}^\mu$, $A_{LR}^\tau$, $A_{LR}^{had}$, 
$A_{LR}^c$, $A_{LR}^b$, $A_{FB}^\mu (pol)$, $A_{FB}^c (pol)$,
$A_{FB}^b (pol)$, and $P_\tau$.  For the $\mu^+\mu^-$ case we included 
the ten observables that did not involve polarized electrons.  In 
calculating the $\chi^2$ we assumed 90\% electron polarization, 35\% 
$c$-tagging efficiency and 60\% $b$-tagging efficiency.  The 99\% C.L. 
discovery limits are shown in figure 3.  Only statistical errors are 
considered in obtaining the limits shown.

Because search limits
obtained at $e^+e^-$ colliders are indirect, based on deviations 
from the standard model in precision measurements, they are sensitive to
both statistical and systematic errors.  If, for example, 
the NLC integrated luminosity is reduced from 50~fb$^{-1}$ to 10~fb$^{-1}$ 
(200~fb$^{-1}$ to 50~fb$^{-1}$) for the 500~GeV (1~TeV) case, the
search limits are reduced by about 33\%.  
Including a 5\% systematic 
error in cross section measurements 
due to, for example, luminosity uncertainties, and a 2\% systematic 
error in asymmetries where systematic errors partially cancel,
typically reduces these numbers by 30\% although in some cases 
(the SSM) it 
can reduce the limits by up to 50\%.  
Clearly, systematic errors will
have to be kept under control for high precision measurements. 
Finally, we did not include 
initial rate radiation (ISR).  Rizzo has found that including ISR can 
lower the search reach by 15--20\% \cite{rizzo}.

One sees that the discovery limits obtained at $e^+e^-$ colliders 
can be quite substantial although they can be quite model dependent.
For example, for the $Z_\psi$,
$C'_L=\pm C'_R$ so that either $C'_V$ or $C'_A=0$.  For
$\sqrt{s}$ sufficiently far away from the $Z^0$ pole deviations are
dominated by $Z^0-Z'$ and $\gamma-Z'$ interference which is proportional
to $C_V^2 {C'_V}^2 +2C_V C_A C'_V C'_A +C_A^2 {C'_A}^2 $.  Since for the
photon $C_A=0$, when $C'_V$ is also equal to 0 deviations from the
standard model become small.

\begin{figure}[h]
\leavevmode
\centerline{\epsfig{file=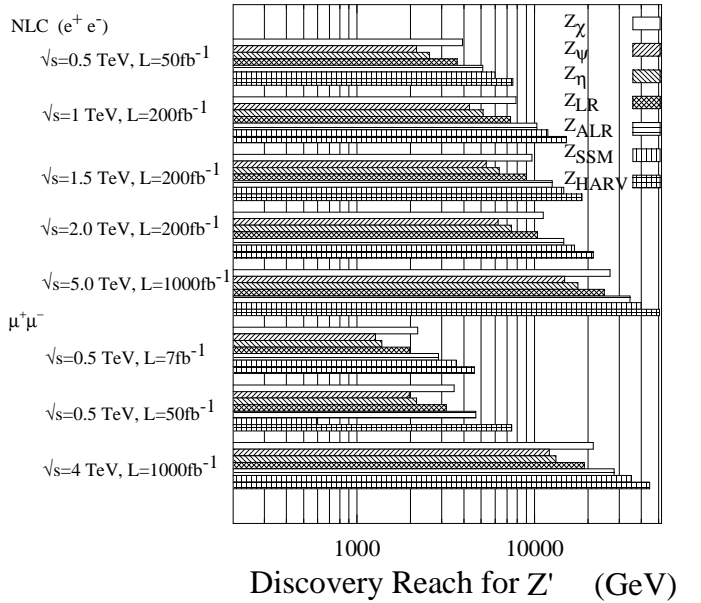,width=9.2cm,clip=}}
\caption{Search limits for extra neutral gauge bosons at high 
energy lepton colliders.  The criteria for obtaining these limits are 
described in the text.}
\end{figure}

\section{Summary}

In this report we have shown that high luminosity lepton colliders can 
put limits on the existence of extra gauge bosons via deviations of 
measurements from their standard model values.  The so called 
discovery limits can be many times the centre of mass energy of the 
collider.  Polarization and flavour tagging are important in realizing 
the highest discovery limits possible.  Since bounds on $Z'$'s are 
based on precision measurements they are quite sensitive to 
measurement errors.  Given the high statistics, to achieve the highest 
possible discovery limits, it will be crucial to minimize systematic 
errors.  Finally, we note that while the non-observation of deviations from 
the SM will put constraints on the existence of $Z'$'s, if deviations 
are found, disentangling the underlying physics will be far from 
trivial.

\section{Acknowledgements}

The author gratefully acknowledges helpful conversations and 
communications with Dean Karlen and JoAnne Hewett and the 
encouragement of Tom Rizzo and Linda Hopson.

%%%%% References
%

\end{document}